\documentclass[
amsmath,amssymb,aps,
]{revtex4-2} 

\usepackage{amsmath}
\usepackage{ulem}
\usepackage{color}
\usepackage{graphicx}
\usepackage{epsfig}
\usepackage{subfig}
\usepackage{bm}
\usepackage{hyperref}
\usepackage{natbib}
\usepackage{pgfplots,mathtools}
\usepackage{hyperref}
\usepackage{amsmath}
\usepackage{braket}
\usepackage{slashed}
\usepackage[compat=1.0.0]{tikz-feynman}
\usepackage{physics}
\usepackage{soul}
\usepackage{xfrac}
\usepackage{algorithm}
\newcommand{\be}{\begin{equation}}
\newcommand{\ee}{\end{equation}}
\newcommand{\bea}{\begin{eqnarray}}
\newcommand{\eea}{\end{eqnarray}}
\newcommand{\ba}[1]{\begin{array}{#1}}
	\newcommand{\ea}{\end{array}}
\newcommand{\nn}{\nonumber}

\newcommand{\ep}{\epsilon}
\newcommand{\om}{\omega} 
\newcommand{\Om}{\Omega}

\begin{document}

\preprint{APS/123-QED}
	\title{Effect of Coriolis force on the shear viscosity of quark matter: A nonrelativistic description}
	\author{Cho Win Aung$^1$, Ashutosh Dwibedi$^1$, Jayanta Dey$^2$\\
		Sabyasachi Ghosh$^1$}

	\affiliation{$^1$Department of Physics, Indian Institute of Technology Bhilai, Kutelabhata, Durg 491001, India}
	\affiliation{$^2$Department of Physics, Indian Institute of Technology Indore, Simrol, Indore 453552, India}

	\begin{abstract}
    Shear viscosity becomes anisotropic in a rotating medium. It is discovered here that for rotating thermalized quantum systems such as those created in relativistic heavy ion collisions, the coefficient of shear viscosity breaks up into five independent components. Similar phenomena were also discovered for quark-gluon plasma in the presence of the magnetic field. Like the Lorentz force at a finite magnetic field, the Coriolis force also creates anisotropic viscosity at nonzero rotation. As a first approach, for simplicity, the calculations are done in the non-relativistic prescription, with a future proposal to extend it toward a relativistic description. 
    Introducing the Coriolis force term in relaxation time approximated Boltzmann transport equation, we have found different effective relaxation times along the parallel, perpendicular, and Hall directions in terms of actual relaxation time and rotating time period. Comparing the present formalism with the finite magnetic field picture, we have shown the equivalence of roles between the rotating and cyclotron time periods, where the rotating time period is inverse of twice the angular velocity.
	\end{abstract}
	\maketitle
	\section{Introduction}
	
	In off-central heavy ion collisions (HIC), a very high orbital angular momentum (OAM) can be deposited. In a typical collision, OAM created from torque at the time of collision could be of the order of $\sim$ ($10^3$ - $10^7$) $\hbar$. The magnitude of this OAM depends on the impact parameter, collision energy, and system size~\cite{STAR:2017ckg, Liang:2004ph, Becattini:2007sr}. A fraction of this initial OAM is transferred to the created quark-gluon plasma (QGP) medium in the form of local vorticity. The impact of such a huge initial OAM or later time vorticity on various observables and polarization has been calculated from various theoretical viewpoints. The Refs.~\cite{becattini2007microcanonical, Becattini:2007sr, BECATTINI20082452, BECATTINI20101566, Becattini:2013fla, Becattini:2013vja, Becattini:2015ska, Becattini:2021suc} have studied the statistical properties with keen interest on the polarization of particles in HIC by demands of angular momentum conservation. Whereas the Refs.~\cite{Betz:2007kg, Liang:2004ph, LIANG200520, Wang_2008,chen2009general, XuGuangHuang2011} have taken the approach of the spin-orbit coupling under strong interactions to explain the polarization observed in HIC. On the other hand, the authors of the Refs.~\cite{Wang_2012, Wang_2013,Wang_2016, Wang_2017,GAO2015542,Yang2017,Gao2017,Gao2018,Liao2018,Gao2019,Gao_2019,Hattori2019,Zhuang_2019M,Rischke2019,yang2020effective,Rischke2021w,Rischke_2021Nk} have taken the approach of quantum kinetic theory to obtain chiral anomalies and polarization effects observed in HIC.
	More recently, a new theoretical framework has been proposed where the complete evolution of spin has been taken care of through explicit incorporation of polarization in a hydrodynamic framework~\cite{Becattini2011,Florkowski2018,Florkowski_2018,Florkowski:2018fap,Florkowski:2018ahw,BECATTINI2019419,Florkowski:2019qdp,BHADURY2021,Bhadury_2021,daher2022equivalence,Bhadury2022}. People have calculated the evolution of vorticity and the polarization of particles with a particular focus on $\Lambda-$hyperon by various transport and hydrodynamical models~\cite{XuGuangHuang2016,Jiang2016,Pang:2016igs,Li:2017slc,Xia:2018tes,XuGuangHuang2019,XuGuangHuang:2019xyr,Wu:2019yiz,XuGuangHuang:2020dtn,Fu:2021pok,XuGuangHuang_2022,XuGuangHuang2022}. There have also been studies on the evolution and thermodynamic properties of QGP and hadronic medium in the presence of rotation~\cite{Sahoo:2023xnu,Pradhan:2023rvf}. See Refs.~\cite{Becattini:2020ngo,Florkowski:2018fap} for recent review papers on the topic related to the vorticity of QGP and polarization of hadrons. 
	The action of Coriolis force in a rotating frame and Lorentz force in an inertial frame are similar.  An analogy was shown between the effect of rotation (Coriolis force) and the magnetic field (Lorentz force) in Refs.~\cite{J_Sivardiere_1983,Johnson2000-px}.  Both magnetic fields and rotation can be produced in the peripheral HIC. Thus, one can observe similar effects on the medium in the presence of rotation as one observes in the presence of magnetic fields. 
	Now, the medium constituents of HIC (quarks and hadrons) have two fundamental quantities, momentum, and spin, which will be affected by both angular velocity and the magnetic field. Momentum will be affected directly via the Lorentz force and Coriolis force. However, spin gets modified through different mechanisms, which is a primary interest to the spin-hydrodynamics community~\cite{Becattini:2020ngo,Florkowski:2018fap}. The spin evolution in the medium is ultimately connected with the experimental quantity - polarization of hadrons. The present article focuses only on the former quantity, momentum, which will be affected by the angular velocity of the medium through the Coriolis force. Our future aim will be to go for a more realistic picture by considering other ingredients like the effect of different (pseudo) forces due to rotation, the impact of angular velocity on the spin, etc.

    Before going to address our work, we have distinct knowledge of three physical quantities connected with the medium rotation: 
    ($1$) local vorticity ($\Vec{\Om}_{l}$), ($2$) global vorticity ($\Vec{\Om}_{g}$), and ($3$) angular momentum density~($\Vec{l}$), which are frequently discussed
     in the literature of rotating QGP topics. They can be briefly defined as follows.
    \begin{itemize}
        \item Local vorticity ($\Vec{\Om}_{l}$): This is defined as $\Vec{\Om}_{l}\equiv\frac{1}{2}\Vec{\nabla}\times \Vec{u}$, where $\Vec{u}$ is the fluid velocity. This quantifies the amount of local rotation or circulation around a loop of the velocity field $\Vec{u}$, similar to the curl of an electric field, which quantifies the circulation of an electric field in electromagnetism.  
        \item Global vorticity ($\Vec{\Om}_{g}$): This corresponds to a situation where fluid as a whole rotates rigidly with angular velocity $\Vec{\Om}$, which gives rise to fluid velocity $\Vec{u}=\Vec{\Om}\times\Vec{r}$. The local vorticity for a fluid evaluated in such a situation merges with $\vec{\Om}$, i.e., $\Vec{\Om}_{l}=\frac{1}{2}\Vec{\nabla}\times \Vec{u}=\frac{1}{2}\Vec{\nabla}\times (\Vec{\Om}\times\Vec{r})=\Vec{\Om}$. Therefore, in this particular case, the fluid is said to have a global vorticity $\Vec{\Om}_{g}\equiv \Vec{\Om}$.
        \item Angular momentum density ($\Vec{l}$): Angular momentum density for a fluid may be defined as $\Vec{l}=\rho (\Vec{r}\times\Vec{u})$, where $\rho$ is the mass density. For example, the angular momentum density for a rigidly rotating fluid with global vorticity $\Vec{\Om_{g}}$ is $\Vec{l}=\rho~\Vec{r}\times(\Vec{\Om}_{g}\times\Vec{r})\implies l_{i}=\rho(|\vec{r}|^{2}\delta_{ij}-x_{i}x_{j})~\Om_{gj}=I_{ij}\Om_{gj}$, where $I_{ij}\equiv\rho(|\Vec{r}|^{2}\delta_{ij}-x_{i}x_{j})$ is the moment of inertia density. However, one may generally have angular momentum density without local vorticity and vice-versa.
 \end{itemize}
 
    In this paper, to simplify the analysis, we will suppose that the particles have an additional random part of the velocity $\Vec{v}$ on top of a rigid rotation velocity $\Vec{\Om}\times \Vec{r}$. This seems to us as a simple way of incorporating angular momentum density into a system by replacing more realistic situations. Adopting this background of including angular momentum into the system, the present paper is planned to concentrate only on the topic of the effect of Coriolis force on the shear viscosity Ref.~\cite{PhysRevC.109.034914}. To fulfill this purpose, we will use the Boltzmann Transport Equation (BTE)-based kinetic theory as our microscopic model. The BTE will be written in the frame rotating with angular velocity $\Vec{\Om}$ to include our background choice implicitly. Again, for simplicity, we will start with the non-relativistic matter with the future aim to extend it towards a relativistic description.

    Recently, Refs.~\cite{Tuchin:2011jw,Ghosh:2018cxb,Mohanty:2018eja,Dey:2019vkn,Dash:2020vxk,Dey:2019axu,Ghosh:2020wqx} have gone through a systematic and step-by-step study on the problem of the effect of Lorentz force on the shear viscosity of magnetized matter. Connecting the similarity between Lorentz force and Coriolis force, the effect of Coriolis force on the shear viscosity of rotating matter is explored here. In the absence of magnetic fields, there is only one velocity gradient term. As a result, the shear viscosity of the medium is isotropic. At a finite magnetic field, the shear stress tensor breaks up into five independent components as one can build five independent velocity gradient tensors in terms of fluid velocity $u_i$ and magnetic field unit vector $b_i$. 
   Similarly, the viscous stress tensors can have five independent velocity gradient components for a fluid under finite rotation in terms of fluid velocity $u^i$ and angular velocity unit vector $\omega_i$. We have developed a detailed formalism to calculate the shear viscosity in the presence of rotation (detailed in Sec.~\ref{sec:form}). Consequently, in Sec.~\ref{sec:res}, we have described the numerical outcomes on temperature and angular velocity dependency of shear viscosity with graphical visualization and interpretation. Ultimately, we have summarized our findings in Sec.~\ref{sec:sum}.
 

	%
	\section{Formalism}
    \label{sec:form}
	\begin{figure}
		\includegraphics [scale=0.5]{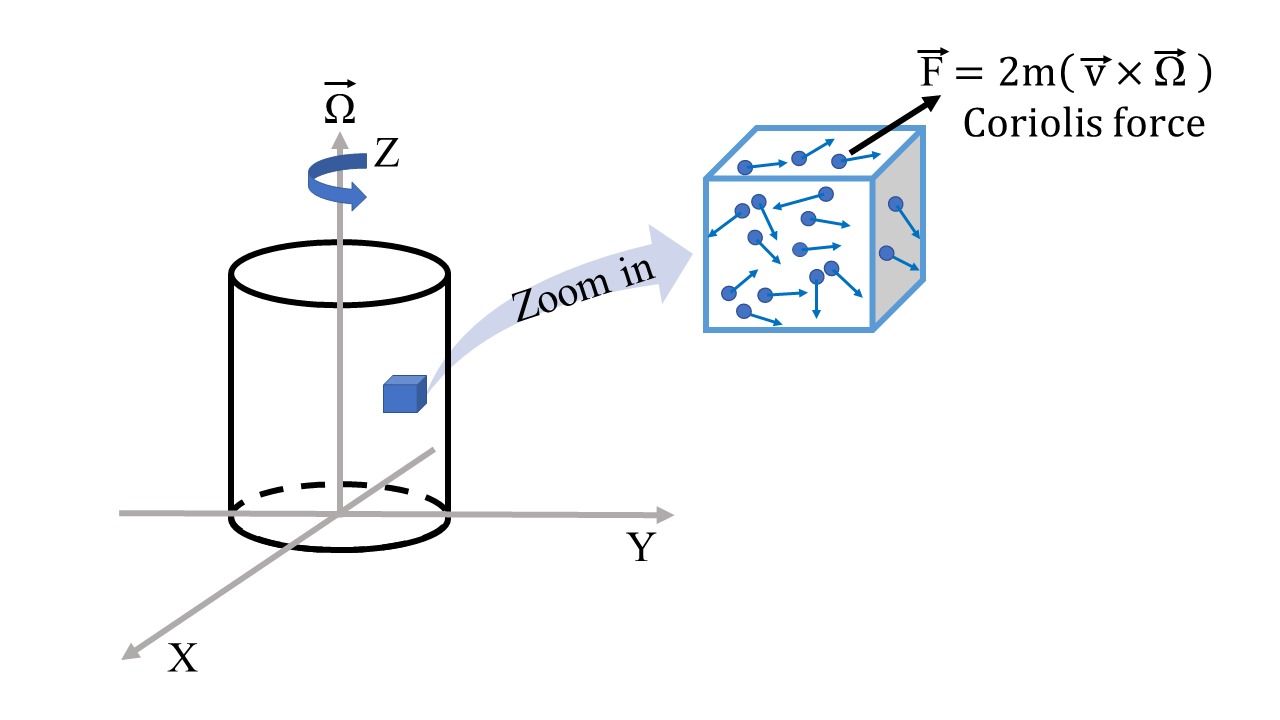}
		\caption{Schematic picture of rotating cylinder with fluid on the left side, whose one of the (cubical) fluid elements is zooming
			in the right side, where particles inside the fluid element box are randomly moving and facing Coriolis force}
		\label{f1}
	\end{figure}
	In classical mechanics, if we have a system rotating with an angular velocity $\Vec{\Om}$, one can write the following operator equation holding for any arbitrary vector~\cite{goldstein2011classical},
	\begin{equation}
	\left(\frac{d}{dt}\right)_{s} \equiv \left(\frac{d}{dt}\right)_{r}+\Vec{\Om}\times~,
	\label{B1}
	\end{equation}
	where $s$ and $r$ in the subscripts of the expression mean, the time-derivative of a vector has to be performed with respect to space-fixed and rotating frames, respectively. If one substitutes the position vector $\vec{r}$ in the operator equations one gets the relation, $\Vec{v}_{s}=\Vec{v}_{r}+\Vec{\Om}\times\vec{r}$, where one identifies $\Vec{v}_{s} $ and $\Vec{v}_{r}$ with velocity in space-fixed and rotating frame respectively. Again substituting this in general Eq.~(\ref{B1}) we have
	\begin{equation}
	\Vec{a}_{s}= \Vec{a}_{r}+2(\Vec{\Om} \times \Vec{v}_{r})+ \Vec{\Om} \times (\Vec{\Om}\times\Vec{r})+ \Dot{\Vec{\Om}} \times \Vec{r}.
    \label{B2}
	\end{equation}
	We will ignore the subscripts $s$ and $r$ on the vectors for simplicity of notation, so, from now onwards, we will call $\Vec{v}_{r}$  and its component as $\vec{v}$ and $ v_{i} $ respectively.
	The terms of Eq.~(\ref{B2}) can be rearranged to write Newton's equation in a rotating frame. The second term in Eq.~(\ref{B2}) is known as the Coriolis acceleration. In Fig.~\ref{f1}, we have schematically presented a cylinder containing fluid rotating with angular velocity~$\Vec{\Om}$. For simple visualization, the geometry of the fluid system has been chosen as cylindrical. If we take any fluid element and look at it closely, the particles inside it have a random part of the velocity $\Vec{v}$ on top of the rotational velocity $\Vec{\Om}\times\Vec{r}$. All the particles inside any fluid element feel the Coriolis force $2m(\Vec{v}\times\Vec{\Om})$. For the case of constant angular velocity (as is assumed here), the Euler force vanishes, but the other two forces, i.e., Coriolis and Centrifugal, remain non-zero. In the present calculation, we will consider only the effect of Coriolis force on particle motion to keep our expressions simple to understand. In a realistic system, both forces should be considered, but one may ignore the centrifugal force for the particular domain where particle (average) velocity is quite larger than the fluid element's angular velocity (more explicitly $v>> \Omega r/2$).
	We can find a similarity or equivalence between finite magnetic fields and finite rotation pictures. For example, at finite magnetic field ($B$), a particle with charge $q$ and velocity $v$ will face the Lorentz force $\vec{F}=q \vec{v}\times \vec{B}$, while at angular velocity $\Omega$ of medium, a particle with mass $m$ and velocity $v$ will face the Coriolis force $\vec{F}=2m \vec{v}\times \vec{\Omega}$. The dissipative part of the energy-momentum tensor is modified at the microscopic level through the Lorentz force. A similar kind of modification can be expected for the finite rotation case. The similarity between this finite $ B$ and finite $\Om$ in microscopic descriptions inspires us to build a similar kind of macroscopic description.
	Refs.~\cite{Tuchin:2011jw,Ghosh:2018cxb,Mohanty:2018eja,Dey:2019vkn,Dash:2020vxk,Dey:2019axu,Ghosh:2020wqx} have prescribed that macroscopic expressions of dissipative energy-momentum tensor at finite $B$ can be built by the basic tensors - fluid velocity $(u_i)$, Kronecker delta $(\delta_{i j})$, and the component of magnetic field unit vector,  $b_{i}(B_i\equiv B b_{i})$. The same macroscopic structure can be expected in finite rotation by replacing $b_{i}$ by angular velocity unit vector $\omega_{i}(\Om_i\equiv \Om \om_{i})$.
	Following the structure similar to the finite magnetic field case, we can write  viscous stress tensor for finite angular velocity as 
	\begin{equation}
	\tau^{i j}=\eta^{ijkl}  U_{kl},
	\label{A1}
	\end{equation}
	where $U_{kl} $=$ \frac{1}{2} (\frac{\partial u_k}{\partial x_l}+\frac{\partial u_l}{\partial x_k})$ is the velocity gradient and $\eta^{ijkl}$ is the viscosity tensor.
    Here, total energy-momentum tensor ($T^{i j}$) is symmetric in nature, and for having finite contribution of viscosity in the entropy production, velocity gradient term is also symmetric in nature~\cite{groot1980relativistic,HUANG20113075, Denicol2021, Jaiswal:2013vta}. We ignored the antisymmetric part of the gradient $\frac{1}{2} (\frac{\partial u_k}{\partial x_l}-\frac{\partial u_l}{\partial x_k})$ in writing Eq.~(\ref{A1}) because it does not contribute to the viscous stress tensor if one only considers first order deviations of the system from equilibrium.\\
	
    Now we can make seven independent tensor components with the properties that they remain symmetric under the exchange of indices $i\leftrightarrow {j}$ and $k\leftrightarrow{l}$~\cite{HUANG20113075}. These tensor components are given below
	\begin{eqnarray}
	&&\delta_{ik} \delta_{jl}+\delta_{jk} \delta_{il},\nn\\
	&&\delta_{ij} \delta_{kl},\nn\\
	&&\delta_{ik} \omega_j \omega_l +\delta_{jk} \omega_i \omega_l+\delta_{il} \omega_j \omega_k +\delta_{jl} \omega_i \omega_k,\nn\\
	&&\delta_{ij} \omega_k \omega_l +\delta_{kl} \omega_i \omega_j,\nn\\
	&&\omega_i \omega_j \omega_k \omega_l,\nn\\  
	&&\omega_{ik}\delta_{jl}+\omega_{jk}\delta_{il} +\omega_{il}\delta_{jk}+\omega_{jl}\delta_{ik}  
	,\nn\\       
	&&\omega_{ik}\omega_j \omega_l+\omega_{jk} \omega_i \omega_l+\omega_{il} \omega_j \omega_k+\omega_{jl} \omega_i \omega_k,
	\label{A2}
	\end{eqnarray}
	where $\omega_{ij}\equiv\ep_{ijk}\om_k$. We can make seven independent tensors $C^n_{ijkl}, (n=0 \text{ to }6)$ with the help of the linear combination of basis given in Eq.(\ref{A2}). The first five tensors $C^n_{ijkl}, (n=0\text{ to }4)$ when contracted with $U_{kl}$ gives five traceless tensors $C^n_{ij}, (n= 0 \text{ to } 4)$, and the last two tensors, $C^n_{ijkl}, (n=5,6)$ upon contraction, give two non-zero trace tensors $C^n_{ij}, ( n= 5,6)$.
	%
%
    Similar to the structure of five traceless tensors and two non-zero trace tensors for finite magnetic field case~\cite{pitaevskii2017course,Tuchin:2011jw,Ghosh:2018cxb,Mohanty:2018eja,Dey:2019vkn,Dash:2020vxk,Dey:2019axu}, $C^n_{ijkl}$ can be expressed as
\begin{eqnarray}
	C_{ijkl}^0&=&(3\omega_i \omega_j-\delta_{ij})(\omega_k \omega_l-\frac{1}{3}\delta_{kl}),\nn\\
	C_{ijkl}^1&=&\delta_{il} \delta_{jk}+\delta_{jl} \delta_{ik}-\delta_{ij} \delta_{kl}+\delta_{ij} \omega_k \omega_l-\delta_{jl}\omega_i\omega_k\nn\\
	&-&\delta_{jk} \omega_i \omega_l+\delta_{kl} \omega_i \omega_j-\delta_{ik} \omega_j \omega_l-\delta_{il} \omega_j \omega_k+\omega_i \omega_j\omega_k \omega_l,\nn\\
	C_{ijkl}^2&=&\delta_{ik} \omega_j \omega_l+\delta_{il} \omega_j \omega_k+\delta_{jk} \omega_i \omega_l+\delta_{jl} \omega_i \omega_k-4\omega_i \omega_j \omega_k \omega_l,\nn\\
	C_{ijkl}^3&=&\delta_{il} \omega_{jk}+\delta_{jl} \omega_{ik}-\omega_{ik} \omega_j \omega_l-\omega_{jk} \omega_i \omega_l,\nn\\
	C_{ijkl}^4&=&\omega_{ik}\omega_j\omega_l+\omega_{il}\omega_j\omega_k+\omega_{jk}\omega_i \omega_l +\omega_{jl} \omega_i \omega_k,\nn\\ 
	C_{ijkl}^5&=&\delta_{ij}\delta_{kl},\nn\\
	C_{ijkl}^6&=&\delta_{ij}\omega_k\omega_l+\delta_{kl}\omega_{i}\omega_{j},
	\label{A3}
	\end{eqnarray}
    with
	\begin{eqnarray}
	C_{ij}^0 &=& (3\omega_i \omega_j-\delta_{ij})(\omega_k \omega_lU_{kl}-\frac{1}{3} \vec {\nabla} \cdot \vec{u}),\nn\\
	C_{ij}^1 &=& 2U_{ij}+\delta_{ij}U_{kl}\omega_k \omega_l-2U_{ik}\omega_j \omega_k-2U_{jk} \omega_k \omega_i+(\omega_i \omega_j-\delta_{ij}) \vec{\nabla} \cdot \vec{u}+\omega_{i}\omega_{j}\omega_{k}\omega_{l} U_{kl},\nn\\
	C_{ij}^2 &=& 2(U_{ik} \omega_j \omega_k+U_{jk} \omega_i \omega_k-2U_{kl}\omega_i \omega_j \omega_k \omega_l),\nn\\
	C_{ij}^3 &=& U_{ik}\omega_{jk}+U_{jk}\omega_{ik}-U_{kl}\omega_{ik}\omega_j\omega_l-U_{kl}\omega_{jk}\omega_i \omega_l,\nn\\
	C_{ij}^4 &=& 2(U_{kl} \omega_{ik} \omega_j \omega_l+U_{kl} \omega_{jk} \omega_i \omega_l),\nn\\
	C_{ij}^5 &=& \delta_{ij}(\vec{\nabla}\cdot\vec{u}),\nn\\
	C_{ij}^6 &=& \delta_{ij}\omega_k \omega_l U_{kl}+\omega_i\omega_j (\vec{\nabla}\cdot\vec{u}),
	\label{A4}
	\end{eqnarray}
    %
	where $C_{ij}^n= C^n_{ijkl}U_{kl}$.
	The viscous tensor can be written as a combination of seven basis tensors given in Eq.~(\ref{A3}) as
	\begin{eqnarray}
	\eta_{ijkl}&=& \eta_0 C^0_{ijkl}+\eta_1 C^1_{ijkl}+\eta_2 C^2_{ijkl}+\eta_3 C^3_{ijkl}+\eta_4 C^4_{ijkl}\nn\\
    &+&\zeta_0 C^5_{ijkl}+\zeta_1 C^6_{ijkl},
	\label{A5}    
	\end{eqnarray}
    where $\eta_0 \text{ to } \eta_{4}$ are designated as shear viscosities, whereas $\zeta_0\text{, and }\zeta_{1}$ are categorized as bulk viscosities of the medium. From now onwards, we will concentrate on the shear viscosities of the medium, ignoring the bulk part of the viscous stress tensor.

	So, the viscous stress tensor given in Eq.~(\ref{A1}) becomes the shear stress tensor, which can be written as
	\begin{eqnarray}
	\pi_{ij}&=&\eta_n C^n_{ijkl} U^{kl},\nn\\
	&=& \eta_n C^n_{ij}.
	\label{A6}
	\end{eqnarray}
    This Eq.~(\ref{A6}) is the macroscopic expression of shear stress tensor $\pi_{ij}$. For its microscopic expression, 
    we have used the kinetic theory framework, which defines the dissipative part of the stress tensor as
	\begin{equation}
	\pi_{ij}=g\int\frac{d^{3}\vec{p}}{(2\pi)^3}mv_iv_j \delta f, 
	\label{A13}
	\end{equation}
	where $g$ is the degeneracy factor of the medium constituent particle with mass $m$ and velocity $v_i=p_i/m$.
    
    In order to ascertain the nature of $\delta f$, the Boltzmann transport equation (BTE) will be employed
	\begin{equation}
	\vec{v}\cdot\pdv{f}{\Vec{r}}+\vec{F}\cdot\frac{\partial f}{\partial\vec{p}}+\frac{\partial f}{\partial t} =\left(\frac{\partial f}{\partial t}\right)_{coll},
	\label{A7}
	\end{equation}
	where $f$ and $\Vec{F}$ are the non-equilibrium distribution function of the particles and the force acting on the particles, respectively.
	The BTE in relaxation time approximation (RTA) can be written as
	\begin{equation}
	\vec{v}\cdot\pdv{f}{\Vec{r}}+\vec{F}\cdot\frac{\partial f}{\partial\vec{p}}+\frac{\partial f}{\partial t}=-\frac{\delta f}{\tau_{c}},
	\label{A8}
	\end{equation}
	where the system has been assumed to be slightly out of equilibrium. The total distribution function comprises two parts- the part corresponding to local equilibrium $f_0$ and a perturbed part $\delta f$, i.e., $f=f_0+\delta f$. $\tau_c$ is the so-called relaxation time for the system.
	Substituting the expression of Coriolis force in place of $\Vec{F}$ and keeping the terms which are 1st order in $\delta f$
	in the LHS of Eq.~(\ref{A8}) we have
	%
	\begin{align}
	\vec{v}\cdot\pdv{f_0}{\Vec{r}}+2(\Vec{v}\times\Vec{\Omega})\cdot\frac{\partial \delta f}{\partial \Vec{v}}+\frac{\partial f_0}{\partial t} &=&-\frac{\delta f}{\tau_{c}}\nn\\
    \implies -f_0(1-f_0) \Vec{v}\cdot \frac{\partial}{\partial \Vec{r}}\frac{E-\mu(\Vec{r},t)-\Vec{u}(\Vec{r},t)\cdot\Vec{p}}{T(\Vec{r},t)}+ 2(\Vec{v}\times\Vec{\Omega})\cdot \frac{\partial \delta f}{\partial \Vec{v}} -f_0(1-f_0)\frac{\partial}{\partial t}\frac{E-\mu(\Vec{r},t)-\Vec{u}(\Vec{r},t)\cdot\Vec{p}}{T(\Vec{r},t)} &=& -\frac{\delta f}{\tau_{c}},
	\label{A9}
	\end{align}
	%
	where the local equilibrium distribution
	$f_0=\Big[\exp\Big(\frac{E-\mu(\Vec{r},t)-\Vec{u}(\Vec{r},t)\cdot\Vec{p}}{T(\Vec{r},t)}\Big)+1\Big]^{-1}$ and $\Vec{u}$ is the fluid velocity.
    The space-time gradient of chemical potential ($\mu$) from the LHS of Eq.~(\ref{A9}) can be replaced with the space-time gradients of pressure ($P$) and temperature ($T$) by the use of Gibbs-Duhem relation. Subsequently, the conservation equations of ideal fluid can also be used to eliminate the time derivative of $P$,~$T$, and $\Vec{u}$. Eventually, one is left with only the space gradients of temperature $\frac{\partial T}{\partial x_{i}}$, and fluid velocity $\frac{\partial u_{i}}{\partial x_{j}}$ in the LHS of Eq.~(\ref{A9})~\cite{kremer2010int,scribdLectureNotes}. Similarly, $\delta f$ can be thought to be made up of two parts, i.e., $\delta f=\delta f_{\eta,\zeta}+\delta f_{\kappa}$. The $\delta f_{\eta,\zeta}$ corresponds to corrections of $f_0$ due to bulk and shear stresses in the fluid. It gives rise to shear viscosity $\eta$ and bulk viscosity $\zeta$. The $\delta f_{\kappa}$ corresponds to the correction of $f_0$ due to temperature gradient. It gives rise to thermal conductivity $\kappa$. The gradients that give rise to shear or bulk viscosities ($\frac{\partial u_{i}}{\partial x_{j}}\text{ or } \frac{\partial u_{i}}{\partial x_{i}}$) and the gradients corresponding to thermal conductivity ($\frac{\partial T}{\partial x_{i}}$) are independent. Therefore, each case can be analyzed separately. We will consider only $\delta f_{\eta}$ in the present article even though, due to notational simplicity, we will call it $\delta f$ onwards.
    By only keeping the terms that correspond to stress in the fluid, the LHS of Eq.~(\ref{A9}) can be written as:
	\begin{equation}
	\frac{mv_iv_j}{T}\frac{\partial u_j}{\partial x_i}f_0(1-f_0)+ 2(\Vec{v}\times\Vec{\Omega})\cdot\frac{\partial \delta f}{\partial \Vec{v}}=-\frac{\delta f}{\tau_{c}},
	\label{A10}
	\end{equation}
	where we have followed Einstein's summation convention.
	Using the identity $U_{ij}\equiv\frac{1}{2}\left(\frac{\partial u_j}{\partial x_i} +\frac{\partial u_i}{\partial x_j}\right)$, we can
	express Eq.~(\ref{A10}) as:
	\begin{equation}
	\frac{m}{T}v_iv_jU_{ij}f_0(1-f_0)+2(\Vec{v}\times\Vec{\Omega})\cdot\frac{\partial \delta f}{\partial \Vec{v}}=-\frac{\delta f}{\tau_{c}}.
	\label{A12}
	\end{equation}
	
	 To access $\pi_{ij}$, we need $\delta f$, which can be acquired by solving Eq.~(\ref{A12}). We will guess the solution of Eq.~(\ref{A12}) as:
	\begin{equation}
	\delta f=\sum_{n=0}^{4}C_n C^n_{kl} v_k v_l. 
	\label{A14}
	\end{equation}
We will see in the following calculation that the above-written guess solution in the form of Eq.~(\ref{A14}) actually works. This justifies the validity of Eq.~(\ref{A1}) with only the symmetric part of the velocity gradient $U_{kl}$. As we have anticipated, the validity of the guess solution in the form of Eq.~(\ref{A14}) can be traced back to the approximation we used in obtaining Eq.~(\ref{A9}) from Eq.~(\ref{A8}). But if one considers higher orders of the approximation, the viscous stress tensor can also contain the anti-symmetric part of the gradient \cite{Denicol2021,Jaiswal:2013vta}.\\ 
	The Eq.~(\ref{A12}) can be rewritten as:
	\begin{eqnarray}
	\frac{m}{T}v_iv_jU_{ij}f_0(1-f_0)+2\ep_{ijk}v_j\om_{k}\Om \frac{\partial \delta f}{\partial v_{i}} &=& -\frac{\delta f}{\tau_{c}}\nn\\
	\implies\frac{m}{T}v_iv_jU_{ij}f_0(1-f_0)+\frac{1}{\tau_\Om}\om_{ij}v_j\frac{\partial \delta f}{\partial v_{i}} &=& -\frac{\delta f}{\tau_{c}},
	\label{A15}
	\end{eqnarray}
	where $\tau_\Omega=\frac{1}{2\Omega}$. We will see later that this $\tau_\Omega$ will play the same role as the cyclotron time period $\tau_B=m/qB$ plays on the transport coefficient expressions at finite magnetic field. Now,
	\begin{equation*}
		\frac{\partial \delta f}{\partial v_i} = \frac{\partial}{\partial v_i}\sum_{n=o}^{4}C_nC^n_{kl}v_k v_l. 
	\end{equation*}
	Using this result of Eq.~(\ref{A15}),
	\begin{align}
		&&\frac{m}{T}v_iv_j U_{ij}f_0(1-f_0)+\frac{2}{\tau_\Om}\om_{ij}v_j\sum_{n=o}^{4}C_n C_{ik}^nv_{k} = -\frac{1}{\tau_{c}}\sum_{n=0}^{4} C_n C_{kl}^n v_kv_l\nn\\
		&&\implies\frac{m}{T}v_iv_jU_{ij}f_0(1-f_0) = \sum_{n=0}^{4} C_n \Big(-\frac{2} {\tau_\Om} \om_{ij} v_j v_k C_{ik}^n -\frac{1}{\tau_c} C_{kl}^n v_kv_l \Big),
        \label{A18}
	\end{align}
	where $\om_{ij}v_i v_j=0$.
	The Eq.~(\ref{A18}) can be further simplified explicitly by expressing $C_{ik}^nv_
	jv_k$ and $C_{kl}^nv_kv_l$ in terms of elementary tensor structures. All the  $C_n$'s can be calculated by equating the coefficients of the independent tensor blocks that appeared in Eq.~(\ref{A18}) to zero.
	By equating the coefficients $v_iv_jU_{ij} ,U_{ij}v_jv_k\om_{ik},U_{ij}v_k\om_{j}\om_{ik}(\vec{v}\cdot\vec{\Om})$ and $U_{ij}v_i\om_j(\vec{v}\cdot\vec{\Om})$ which occurs in the Eq.~(\ref{A18}) to zero, the following set of equations can be attained
	\begin{eqnarray}
	v_i v_j U_{ij} &:& -\frac{4C_3}{\tau_{\Om}} -\frac{2C_1}{\tau_c} = \frac{m}{T} f_0(1-f_0),\nn\\
	U_{ij}v_jv_k\om_{ik} &:& -\frac{4C_1}{\tau_{\Om}}+\frac{2C_3}{\tau_c}=0,\nn\\
	U_{ij}v_k\om_{j}\om_{ik}(\vec{v}\cdot\vec{\Om}) &:& \frac{4C_1}{\tau_{\Om}}-\frac{4C_2}{\tau_\Om}-\frac{2C_3}{\tau_{c}}+\frac{4C_4}{\tau_c}=0,\nn\\
	U_{ij}v_i\om_j(\vec{v}\cdot\vec{\Om}) &:& \frac{8C_3}{\tau_{\Om}}-\frac{4C_4}{\tau_\Om}+\frac{4C_1}{\tau_{c}}-\frac{4C_2}{\tau_c}=0.
	\label{A19}
	\end{eqnarray}
	Solving the above set of linear equations, we have
	\begin{eqnarray}
	&&C_1=-\frac{m}{2T}f_0(1-f_0)\frac{\tau_c}{1+4(\tau_c/\tau_\Om)^2},\nn\\
	&&C_2=-\frac{m}{2T}f_0(1-f_0)\frac{\tau_c}{1+(\tau_c/\tau_\Om)^2},\nn\\
	&&C_3=-\frac{m}{T}f_0(1-f_0)\frac{\tau_c(\tau_c/\tau_\Om)}{1+4(\tau_c/\tau_\Om)^2},\nn\\
	&&C_4=-\frac{m}{2T}f_0(1-f_0)\frac{\tau_c(\tau_c/\tau_\Om)}{1+4(\tau_c/\tau_\Om)^2}.
	\label{A20}
	\end{eqnarray}
	Substituting the value of $\delta f$ in Eq.~(\ref{A13}), and using the result $\int v_iv_jv_kv_l~d^{3}\vec{v}=\frac{v^4}{15}(\delta_{ij}\delta_{kl}+\delta_{ik}\delta_{jl}+\delta_{il}\delta_{jk})d^3v, (d^3 v\equiv4\pi v^2dv)$ we have
	\begin{eqnarray}
	\pi_{ij} &=& g\int\frac{d^{3}\vec{p}}{(2\pi)^3}m\sum_{n=0}^{4}C_n C_{kl}^n v_i v_jv_kv_l
    \nn\\
	&=& g\int{d^{3}v}\frac{m^4}{(2\pi)^3} \sum_{n=0}^{4} C_n C_{kl}^n (\delta_{ij} \delta_{kl} +\delta_{ik}\delta_{jl} +\delta_{il} \delta_{jk}) \frac{v^4}{15}\nn\\
	&=& \frac{2 g m^4}{15} \sum_{n=0}^{4} C_{ij}^n \int \frac{d^3v}{(2\pi)^3} v^4 C_n ,
	\label{A21}
	\end{eqnarray}
	where $C_{kl}^n (\delta_{ij} \delta_{kl} +\delta_{ik} \delta_{jl} +\delta_{il}\delta_{jk})=2C_{ij}^n$. Substituting the values of $C$'s from Eq.~(\ref{A20}) in Eq.~(\ref{A21}), we get the corresponding viscosities as
	\begin{equation}
	\eta_n=-\frac{2 g m^4}{15}\int \frac{d^3v}{(2\pi)^3}v^4 C_n.
	\label{A22}
	\end{equation}
	The $\eta_0$ is the viscosity in the absence of rotation, which will be the same as the expression in the absence of magnetic field case; therefore, it is given by~\cite{Dey:2019axu,Dey:2019vkn}
	$$\eta_0=\frac{g\tau_c}{15T}\int\frac{d^3p}{(2\pi)^3}\frac{p^4}{m^2}f_0(1-f_0).$$
	From the Eq.~(\ref{A22}), we get
	\begin{eqnarray}
	&&\eta_1=\frac{g}{15T}\frac{\tau_c}{1+4(\tau_c/\tau_\Om)^2}\int\frac{d^3p}{(2\pi)^3}\frac{p^4}{m^2}f_0(1-f_0),\nn\\
	&&\eta_2=\frac{g}{15T}\frac{\tau_c}{1+(\tau_c/\tau_\Om)^2}\int\frac{d^3p}{(2\pi)^3}\frac{p^4}{m^2}f_0(1-f_0),\nn\\
	&&\eta_3=\frac{2g}{15T}\frac{\tau_c(\tau_c/\tau_\Om)}{1+4(\tau_c/\tau_\Om)^2}\int\frac{d^3p}{(2\pi)^3}\frac{p^4}{m^2}f_0(1-f_0),\nn\\
	&&\eta_4=\frac{g}{15T}\frac{\tau_c(\tau_c/\tau_\Om)}{1+(\tau_c/\tau_\Om)^2}\int\frac{d^3p}{(2\pi)^3}\frac{p^4}{m^2}f_0(1-f_0).
	\label{A23}
	\end{eqnarray}
    
    Comparing the final expressions of $\eta_n$ at finite $\Om$ with the same for finite $B$, addressed in Refs.~\cite{Dey:2019vkn,Dey:2019axu},
    the reader can find the similarities in mathematical structure if he equates $\tau_\Om\equiv\tau_B$, i.e., $\frac{1}{2\Om}\equiv\frac{m}{qB}$, which may be understood
    as an equivalence between Coriolis and Lorentz forces  
    \begin{eqnarray}
        {\vec v}\times 2m{\vec \Om}&\equiv& {\vec v}\times q{\vec B},
        \nn\\
        \Rightarrow 2m\Om&\equiv& qB.
    \end{eqnarray}
    The above expressions of viscosities can be cast in terms of the Fermi function as follows
	\begin{eqnarray}
	\int_{0}^{\infty}dp~p^6 f_0(1-f_0)&=& \int_{0}^{\infty}dp~ p^6 \left(T\frac{\partial f_0}{\partial \mu}\right)\nn\\
	&=& T\frac{\partial}{\partial \mu}\int_{0}^{\infty}dpf_0 p^6\nn\\
	&=& 4\sqrt{2}Tm^{7/2}\frac{\partial}{\partial \mu}\int_{0}^{\infty}dE f_0 E^{5/2}\nn\\
	&=&  4\sqrt{2}Tm^{7/2}\frac{\partial}{\partial\mu}\int\frac{E^{(7/2)-1}{dE}}{e^{(E-\mu)/T}+1}\nn\\
	&=& 4\sqrt{2}T^{7/2}m^{7/2} \Big(T\frac{\partial}{\partial\mu} \int \frac{x^{(7/2)-1}dx}{A^{-1}e^{x}+1}\Big),
	\label{A24}
	\end{eqnarray}
	where $x=E/T$ and $A=e^{\mu/T}.$ The Fermi function is defined as $f_j(A)\equiv\frac{1}{\Gamma(j)}\int_{0}^{\infty}\frac{x^{j-1}}{A^{-1}e^{x}+1}~dx$, with the property that $\frac{\partial}{\partial (\mu/T)}f_j(A)=f_{j-1}(A)$.
	Using the above definition, we have
	\begin{align}
		\int_{0}^{\infty}dp~ p^6f_0(1-f_0)&=\frac{15}{2}\sqrt{2\pi}m^{7/2}f_{5/2}(A)T^{7/2}.
		\label{A25}
	\end{align}
	Using the result of Eq.~(\ref{A25}) in Eq.~(\ref{A24}) we have
	\begin{align}
		\eta_1&=g\left(\frac{m}{2\pi}\right)^{3/2}\frac{\tau_c}{1+4(\tau_c/\tau_\Om)^2} T^{5/2}f_{5/2}(A),\nn\\
		\eta_2&=g\left(\frac{m}{2\pi}\right)^{3/2}\frac{\tau_c}{1+(\tau_c/\tau_\Om)^2} T^{5/2}f_{5/2}(A),\nn\\
		\eta_3&=g\left(\frac{m}{2\pi}\right)^{3/2}\frac{\tau_c(2\tau_c/\tau_\Om)}{1+4(\tau_c/\tau_\Om)^2} T^{5/2}f_{5/2}(A),\nn\\
		\eta_4&=g\left(\frac{m}{2\pi}\right)^{3/2}\frac{\tau_c(\tau_c/\tau_\Om)}{1+(\tau_c/\tau_\Om)^2} T^{5/2}f_{5/2}(A).  
		\label{A26} 
	\end{align}
	Following the similarity in the definition of parallel, perpendicular, and Hall shear viscosity 
	components $\eta_{\parallel,\perp,\times}$ at finite magnetic field~\cite{Dey:2019axu,Dey:2019vkn}, one can define $\eta_{\parallel}=\eta_1$,
	$\eta_{\perp}=\eta_2$, $\eta_{\times}=\eta_4$.
	\section{Results}
    \label{sec:res}
	\begin{figure}
		\begin{center}
		    \includegraphics[scale=0.4]{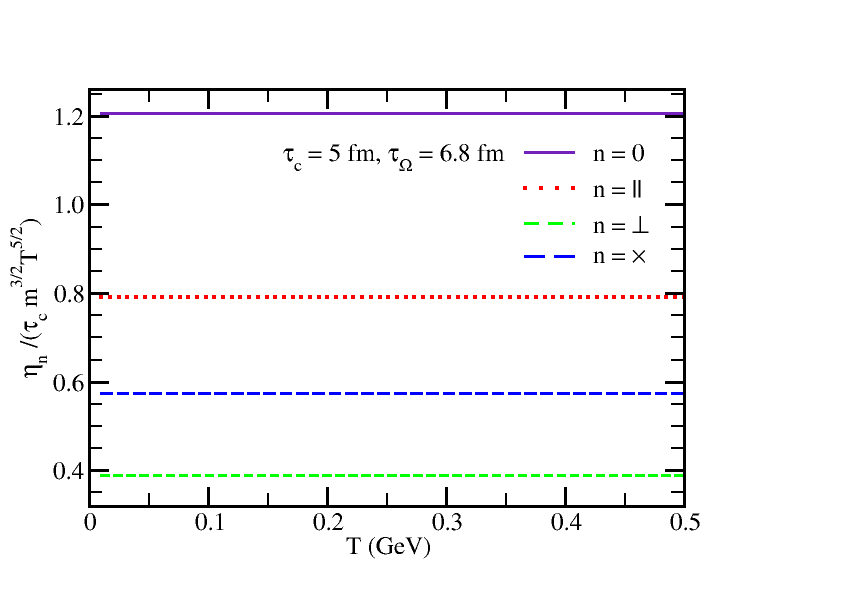}
		    \caption{Normalized parallel ($n=\parallel$), perpendicular ($n=\perp$), and Hall ($n=\times$) components of shear viscosity as well as shear viscosity without rotation are plotted against temperature. }
		\label{fig:fig2}
		\end{center}
	\end{figure}  
	In Sec.~(\ref{sec:form}), we got general expressions of different shear viscosity components for non-relativistic fermionic matter, which can apply to any temperature values $(T)$, chemical potential $(\mu)$ and angular velocity $(\Omega)$. 
	One may readily apply the expression for non-relativistic fluid, belonging to the subject of condensed matter physics 
	and mechanical engineering, where the quantities $T$, $\mu$, and $\Omega$ will be the order of eV in the natural unit.
	However, our destined system belongs to the subject of high-energy nuclear physics and astrophysics, where MeV will be
	the order of magnitude for the quantities $T$, $\mu$, and $\Omega$. Imagining the quark-hadron phase transition
	$T-\mu$ diagram, we can expect two extreme domains - (1) the early universe scenario of net quark/baryon-free domain
	(i.e., at $\mu=0$), which can be produced in LHC and RHIC experiments, and (2) the compact star scenario of 
	degenerate electron or neutron or quark matter (i.e., at $T=0$), expected in white dwarfs and neutron stars. Our microscopic
	expressions of shear viscosity components at finite rotation can be easily applicable to RHIC/LHC matter by putting
	$\mu=0$ and to compact star by putting $T=0$ in the general forms of Eq.~(\ref{A26}). However, we have limitations
	for using non-relativistic matter, which can provide some overestimation with respect to the actual relativistic matter
	expected in RHIC/LHC experiments and compact stars. Our future goal is to reach that actual scenario by developing
	the framework step by step.
	By putting $\mu=0$ and $A=e^{\mu/T}=1$ in Eq.~(\ref{A26}), we get
	\begin{align}
		\eta_{\parallel}=\eta_1&=0.64g\left(\frac{m}{2\pi}\right)^{3/2}\frac{\tau_c}{1+4(\tau_c/\tau_\Om)^2} T^{5/2}\zeta(5/2),\nn\\
		\eta_{\perp}=\eta_2&=0.64g\left(\frac{m}{2\pi}\right)^{3/2}\frac{\tau_c}{1+(\tau_c/\tau_\Om)^2} T^{5/2}\zeta(5/2),\nn\\
		\eta_{\times}=\eta_4&=0.64g\left(\frac{m}{2\pi}\right)^{3/2}\frac{\tau_c(\tau_c/\tau_\Om)}{1+(\tau_c/\tau_\Om)^2} T^{5/2}\zeta(5/2),  
		\label{A26_mu0} 
	\end{align}
	as Fermi function become $f_{5/2}(A=1)=(1-\frac{1}{2^{3/2}})\zeta(5/2)$.   
	Using Eq.~(\ref{A26_mu0}), we have plotted $\eta_{||, \perp,\times}/\tau_{c}m^{3/2}T^{5/2}$ against $T$-axis in Fig.~\ref{fig:fig2}
	and we get horizontal lines as all components are proportional to $T^{5/2}$. We consider quark matter with mass, $m = 0.005$ GeV and
	relaxation time $\tau_c =5$ fm and angular time period $\tau_{\Omega}= 35$ GeV$^{-1}= 6.8$ fm for angular 
	velocity $\Omega=\frac{1}{2\tau_{\Omega}}=0.014$ GeV. We keep comparable values of two-time scales, for which we can get a noticeable difference between parallel and perpendicular components of shear viscosity. 
	\begin{figure}
		\centering
		\includegraphics[scale=0.4]{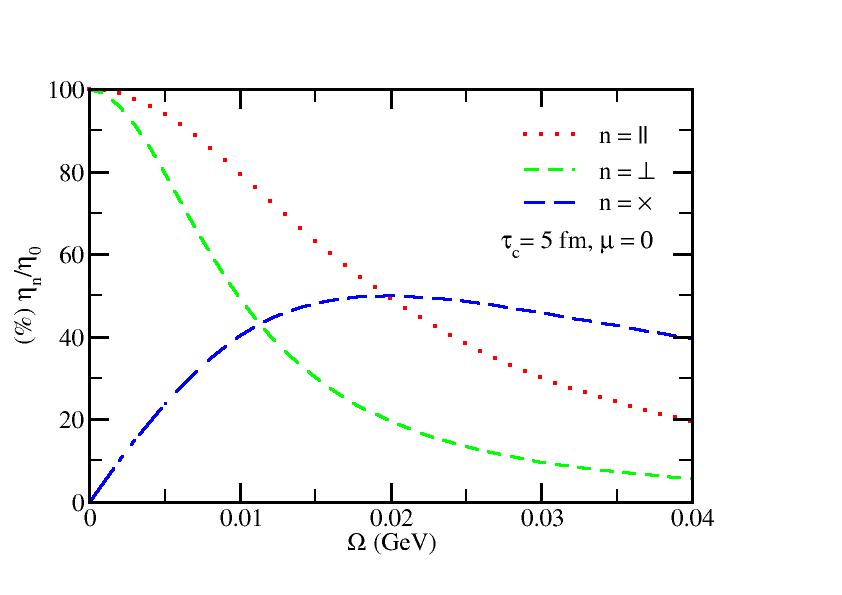}
		\caption{Relative percentage of parallel ($n=\parallel$), perpendicular ($n=\perp$), Hall ($n=\times$) components of shear viscosity vs angular velocity.}
		\label{fig:fig3}
	\end{figure} 
	We can understand the $\eta_{\parallel,\perp,\times}$ in terms of effective relaxation time,
	\begin{align}
		\tau_{\parallel}&=\frac{\tau_c}{1+4(\tau_c/\tau_\Om)^2},\nn\\
		\tau_{\perp}&=\frac{\tau_c}{1+(\tau_c/\tau_\Om)^2}, \nn\\
		\tau_{\times}&=\frac{\tau_c(\tau_c/\tau_\Om)}{1+(\tau_c/\tau_\Om)^2},  
		\label{tau_eff} 
	\end{align}
	as $\eta_{\parallel,\perp,\times}\propto \tau_{\parallel,\perp,\times}$, while $\eta_0\propto \tau_c$ only.
	So we can easily understand that the non-zero ratio $\tau_c/\tau_\Omega$ for finite rotation will
	create the inequality $\tau_{\parallel,\perp,\times}<\tau_c$ and the ratio is also the deciding factor 
	for the ranking among $\eta_\parallel$, $\eta_\perp$, $\eta_\times$. In Fig.~\ref{fig:fig2}, for present
	set of parameters $\tau_c =5$ fm, $\tau_{\Omega}= 6.8$ fm and ratio $\tau_c/\tau_\Omega=0.73$, we get the
	ranking $\eta_\parallel > \eta_\times > \eta_\perp$ but it can be changed for different values of the ratio
	$\tau_c/\tau_\Omega$. This fact will be more clear in the next plot.
	
	In Fig.~\ref{fig:fig3}, we have plotted the percentage of normalized viscosities ($\eta_n/\eta_0$) with respect to $\Omega$ at $\tau_c =5$ fm. It is clearly seen in the plot that the relative magnitude of $\eta_{\perp,||}$ decreases with $\Omega$ in the whole range, whereas $\eta_\times$ initially increases and then decreases with $\Omega$. In the lower range of $\Omega $, $\eta_{\perp,||}$ are more dominant than $\eta_\times$, on contrary in higher range of $\Omega$,  $\eta_\times$ is more dominant than $\eta_{\perp,||}$.
	One can identify both $\eta_{\perp,||}$ will merge to $\eta_0$ in the absence of angular velocity, i.e, $\eta_{\perp,||}(\Omega\xrightarrow{}0)=\eta_0$. From this fact, we can conclude that the finite angular velocity can create anisotropy in shear viscosity components, as we have noticed in the finite magnetic field picture.
	
\begin{figure}
		\centering
		\includegraphics[scale= 0.22]{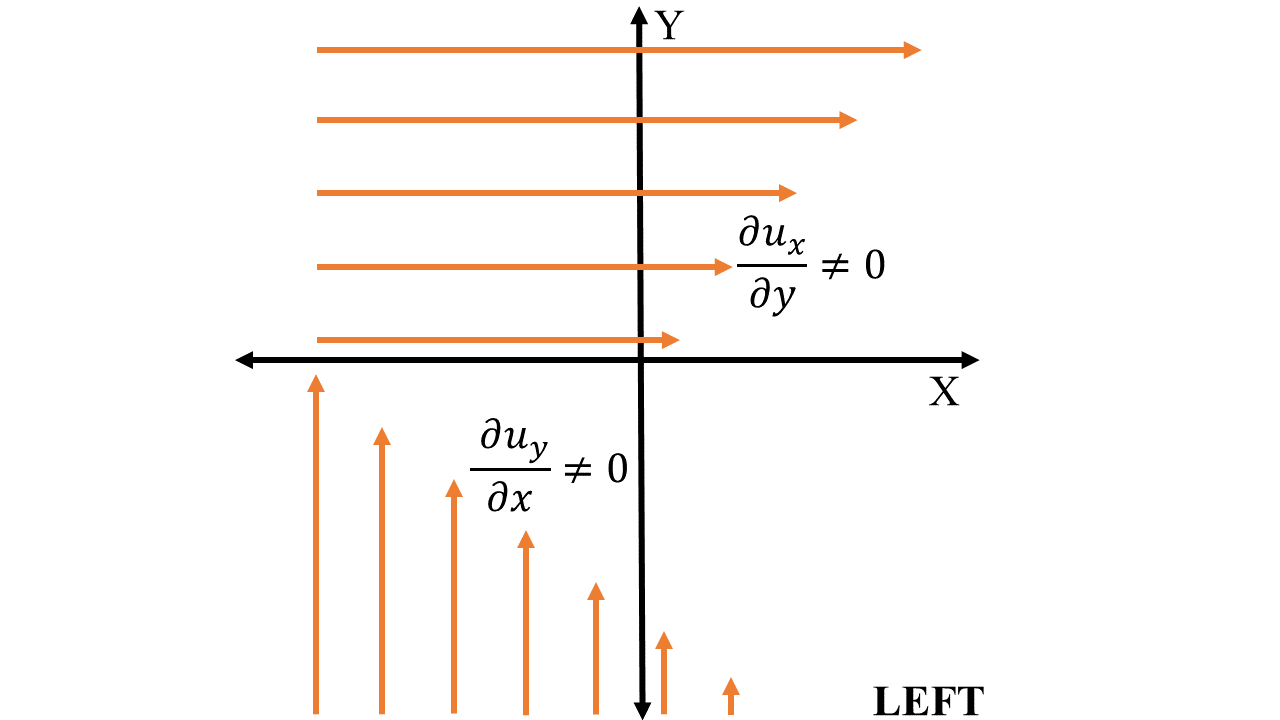}
		\includegraphics[scale= 0.22]{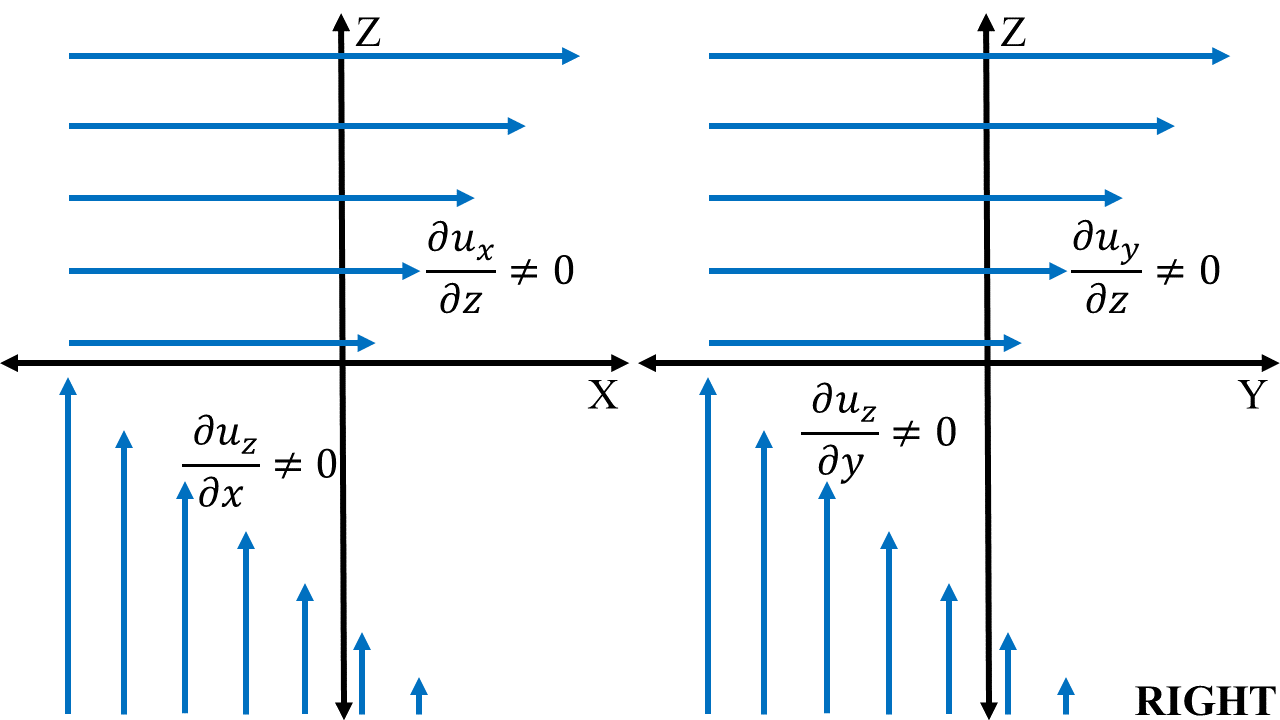}
		\caption{Velocity gradients along XY, ZX, and ZY plane}
		\label{f2}
	\end{figure}

	Let us visualize the different shear viscosity components by means of a schematic diagram - Fig.~\ref{f2}. The picture
	resembles the finite magnetic field picture described in Ref.~\cite{Hattori:2022hyo}. Only the direction of the magnetic field
	along the z-direction will be replaced by the direction of angular velocity. 
	In Fig.~\ref{f2}, the arrows represent the velocity direction, and their lengths represent their order of magnitudes, so changing the arrow lengths map the velocity gradient picture. The right and left panels of  Fig.~\ref{f2} represent the gradient of velocity in the planes, which are parallel (ZX and ZY plane) and perpendicular (XY plane) to the angular velocity, respectively. 
	%
	
\begin{figure}
		\centering
		\includegraphics[scale=0.4]{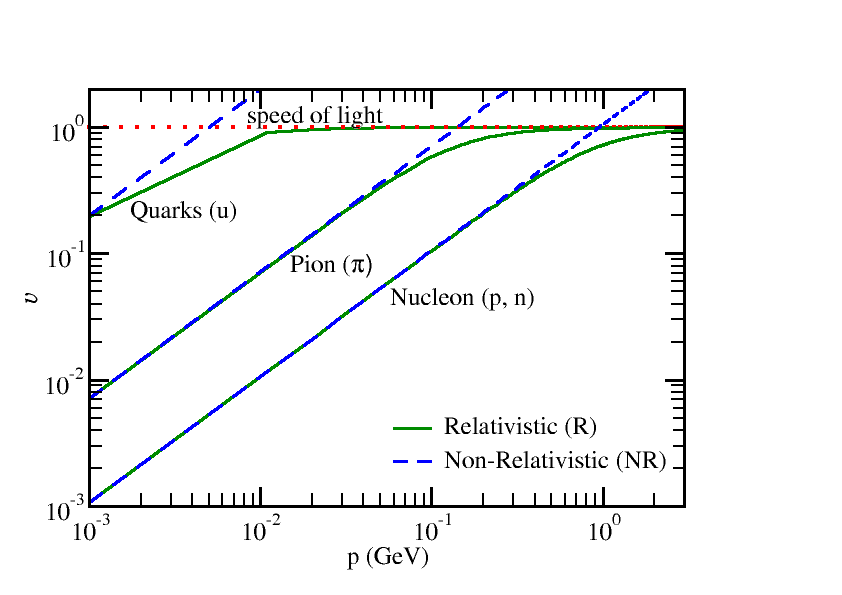}
		\caption{Velocity ($v$) vs momentum ($p$) relation for u quark, $\pi$ meson and nucleons.}
		\label{fig:v_p}
	\end{figure}

	Apart from the rotating quark matter system at $\mu=0$, we can apply the microscopic expressions in Eq.~(\ref{A26_mu0})
	for rotating hadronic matter at $\mu=0$, although the magnitude of angular momentum will be reduced to a smaller value
	in hadronic phase expansion. Considering the $T\rightarrow 0$ limit of Eq.~(\ref{A26}), we can get
	\begin{align}
		\eta_{\parallel}=\eta_1&=\frac{8g}{15\sqrt{\pi}}\left(\frac{m}{2\pi}\right)^{3/2}\frac{\tau_c}{1+4(\tau_c/\tau_\Om)^2} \mu^{5/2}\nn\\
		\eta_{\perp}=\eta_2&=\frac{8g}{15\sqrt{\pi}}\left(\frac{m}{2\pi}\right)^{3/2}\frac{\tau_c}{1+(\tau_c/\tau_\Om)^2} \mu^{5/2}\nn\\
		\eta_{\times}=\eta_4&=\frac{8g}{15\sqrt{\pi}}\left(\frac{m}{2\pi}\right)^{3/2}\frac{\tau_c(\tau_c/\tau_\Om)}{1+(\tau_c/\tau_\Om)^2} \mu^{5/2},  
		\label{A26_T0} 
	\end{align}
	which may be applicable for rotating compact star systems like white dwarfs, neutron stars, and quark matter (expected
	in the core of a neutron star). However, an over-estimation of shear viscosity components of those rotating media
	can be expected by considering the non-relativistic description of relativistic matter. This fact can be understood from 
	the Fig.~\ref{fig:v_p}, where the relativistic and non-relativistic velocity $(v)$ of u quark, pion, and nucleon are plotted
	against momentum $(p)$. From this simple picture, one can see the noticeable difference between relativistic (R) and non-relativistic (NR)
	curves beyond the threshold momenta $1$ MeV, $30$ MeV and $300$ MeV for u quark, $\pi$ meson and nucleon respectively. Overestimation in NR description 
	with respect to R description, will come in the momentum integral beyond those threshold values. Our future aim is to go for
	that relativistic description with an appropriate relativistic extension of the present framework.

    Regarding the fluidity of the medium, quantified by shear viscosity to entropy density ratio, we can find a possibility of violation of KSS bound~\cite{Kovtun:2004de} due to rotation of medium through Coriolis force just like finite magnetic field picture by means of Lorentz force. The entropy density of non-relativistic matter in two extreme limits follow the relations - $s\propto T^{3/2}$ at $\mu\rightarrow 0$ and $s\propto \mu^{3/2}$ at $T\rightarrow 0$. The ratio between shear viscosity to entropy density will be $\eta/s=\frac{\tau_c T}{5}$ at $\mu\rightarrow 0$ and $\eta/s=\frac{\tau_c \mu}{5}$ at $T\rightarrow 0$, which can reach to the KSS bound $\frac{1}{4\pi}$~\cite{Kovtun:2004de} for relaxation time $\tau_c(T)=\frac{5}{4\pi T}$ and $\tau_c(\mu)=\frac{5}{4\pi \mu}$ respectively. At finite rotation, we can expect lower limit expressions for parallel, perpendicular, and Hall components of shear viscosity to entropy density ratio as,
 \bea
    \frac{\eta_{\parallel}}{s}&=&\frac{1}{4\pi}\frac{1}{1+4\Big(\frac{5}{4\pi T\tau_\Om}\Big)^2}\nn\\
    \frac{\eta_{\perp}}{s}&=&\frac{1}{4\pi}\frac{1}{1+\Big(\frac{5}{4\pi T\tau_\Om}\Big)^2}\nn\\
    \frac{\eta_{\times}}{s}&=&\frac{1}{4\pi}\frac{\Big(\frac{5}{4\pi T\tau_\Om}\Big)}{1+\Big(\frac{5}{4\pi T\tau_\Om}\Big)^2}.  
\label{eta_s} 
 \eea
    The above expressions are for $\mu=0$. By replacing $T$ by $\mu$ in Eq.~(\ref{eta_s}), one can get their corresponding expression for $T=0$. So, one can notice that by increasing angular velocity or decreasing $\tau_{\Omega}$ of the medium, $\eta_{\parallel,\perp}/s$ can go below $\frac{1}{4\pi}$. The $\eta_\parallel/s<1/(4\pi)$ is also expected and pointed out by Ref.~\cite{Critelli:2014kra} for finite magnetic field. As a matter of fact, a quantum version extension of the present formalism may be required to comment something on the lower bounds of $\eta_{\parallel,\perp}/s$.
	\section{Summary}
    \label{sec:sum}
	In summary, we have explored the equivalence role of magnetic field and rotation on shear viscosity using Lorentz force and Coriolis force, respectively. In the absence of magnetic fields or rotation, we get an isotropic shear viscosity coefficient, which is proportional to relaxation time only.
    Meanwhile, at a finite magnetic field or rotation, we get anisotropic shear viscosity coefficients proportional to effective relaxation time along the parallel, perpendicular, and Hall directions. This effective relaxation time can be expressed in terms of actual relaxation time and cyclotron-type time period due to magnetic field or rotation. The physics and mathematical steps of the microscopic calculation of shear viscosity at a finite magnetic field or rotation are similar. 
    The fluid velocity gradient is a macroscopic quantity that leads to deviating the total single-particle distribution function out of equilibrium, which is a microscopic quantity. The solution of the Boltzmann equation yields the microscopic expression of the shear stress tensor. Finally, we obtained different components of shear viscosity coefficients in terms of microscopic variables by comparing the microscopic expression with the hydrodynamic expression of the shear stress tensor, a macroscopic expression. 
    This approach obtained anisotropic (or isotropic) shear viscosity components in the presence (or absence) of rotation or magnetic field. At a finite magnetic field, anisotropy is introduced by the Lorentz force term in the Boltzmann equation, and for the finite rotation case, it is the Coriolis force. 
    The present article has only explored the detailed calculation of the finite rotation case. During the description, we have also mentioned the equivalence with the finite magnetic field case. For simplicity, we have attempted it for non-relativistic matters. However, our immediate plan is to extend it toward a relativistic description. To our knowledge, it is the first time we have addressed this anisotropic structure of shear viscosity of rotating matter due to the Coriolis force. 
    We have noticed an equivalence role between the rotating time period for the finite rotation case and the cyclotron time period for the finite magnetic field case, where the rotating time period is defined as the inverse of twice the angular velocity. The factor two propagates from the basic definition of the Coriolis force.
	
	\section*{Acknowledgements}
	CWA acknowledges the DIA programme. This work was partially supported by the Doctoral fellowship in India (DIA) programme of the Ministry of Education, Government of India. AD gratefully acknowledges the Ministry of Education, Government of India. JD gratefully acknowledges the DAE-DST, Government of India funding under the mega-science project – “Indian participation in the ALICE experiment at CERN” bearing Project No. SR/MF/PS-02/2021- IITI (E-37123). SG  thanks Deeptak Biswas and Arghya Mukherjee for the useful discussion during the beginning stage of the work.

	\bibliography{reference}
\end{document}